\documentclass[preprint,showpacs,letterpaper,preprintnumbers,amsmath,amssymb,prdrc]{revtex4}
\usepackage{natbib}
\usepackage{graphicx}
\newcommand{\ppbar}{$p\overline{p}$~}

\newcommand{\MET}{\mbox{$\raisebox{.3ex}{$\not$}E_T$\hspace*{0.5ex}}}

\def\simge{\mathrel{%
   \rlap{\raise 0.511ex \hbox{$>$}}{\lower 0.511ex \hbox{$\sim$}}}}
\def\simle{\mathrel{
   \rlap{\raise 0.511ex \hbox{$<$}}{\lower 0.511ex \hbox{$\sim$}}}}
\def\lessim{\mathrel {\vcenter {\baselineskip 0pt \kern 0pt
\hbox{$<$} \kern 0pt \hbox{$\sim$} }}}
\def\gessim{\mathrel {\vcenter {\baselineskip 0pt \kern 0pt
\hbox{$>$} \kern 0pt \hbox{$\sim$} }}}
\def \rightdownarrow
 {\kern.4em {\raise1.75ex\hbox{$\Bigl |$}$\kern-0.27em{\longrightarrow}$}}
\def \mrightdownarrow
 {\kern.4em {\raise1.25ex \hbox{$|$} \kern-0.27em{
                                                \longrightarrow}}}

\def\mGeVcc{{\rm GeV}\!/c^2}
\def\mGeVc{{\rm GeV}\!/c}

\def\mMEt{\not\kern-.35em {E_T}}

\def\minvpb{~{\rm pb^{-1}}}

\def\simlt{\,^{<}\!\!\!\!{_{\sim}}\,}

\font\eightit=cmti8
\def\r#1{\ignorespaces $^{#1}$}
\begin{document}


\widetext{
\begin{Large}
\begin{center}
Measurement of the Forward-Backward Charge Asymmetry from $W \rightarrow e\nu$ Production in 
\ppbar Collisions at $\sqrt{s} = 1.96$ TeV
\end{center}
\end{Large}

\font\eightit=cmti8
\def\r#1{\ignorespaces $^{#1}$}
\hfilneg
\begin{sloppypar}
\noindent 
D.~Acosta,\r {16} J.~Adelman,\r {12} T.~Affolder,\r 9 T.~Akimoto,\r {54}
M.G.~Albrow,\r {15} D.~Ambrose,\r {43} S.~Amerio,\r {42}  
D.~Amidei,\r {33} A.~Anastassov,\r {50} K.~Anikeev,\r {15}
% ...
\end{sloppypar}
\vskip .026in
\begin{center}
(CDF Collaboration)
\end{center}

\vskip .026in
\begin{center}
\r 1  {\eightit Institute of Physics, Academia Sinica, Taipei, Taiwan 11529, 
Republic of China} \\
\r 2  {\eightit Argonne National Laboratory, Argonne, Illinois 60439} \\
\r 3  {\eightit Institut de Fisica d'Altes Energies, Universitat Autonoma
de Barcelona, E-08193, Bellaterra (Barcelona), Spain} \\
\r 4  {\eightit Istituto Nazionale di Fisica Nucleare, University of Bologna,
I-40127 Bologna, Italy} \\
\r 5  {\eightit Brandeis University, Waltham, Massachusetts 02254} \\
\r 6  {\eightit University of California at Davis, Davis, California  95616} \\
\r 7  {\eightit University of California at Los Angeles, Los 
Angeles, California  90024} \\
\r 8  {\eightit University of California at San Diego, La Jolla, California  92093} \\ 
\r 9  {\eightit University of California at Santa Barbara, Santa Barbara, California 
93106} \\ 
\r {10} {\eightit Instituto de Fisica de Cantabria, CSIC-University of Cantabria, 
39005 Santander, Spain} \\
\r {11} {\eightit Carnegie Mellon University, Pittsburgh, PA  15213} \\
\r {12} {\eightit Enrico Fermi Institute, University of Chicago, Chicago, 
Illinois 60637} \\
\r {13}  {\eightit Joint Institute for Nuclear Research, RU-141980 Dubna, Russia}
\\
\r {14} {\eightit Duke University, Durham, North Carolina  27708} \\
\r {15} {\eightit Fermi National Accelerator Laboratory, Batavia, Illinois 
60510} \\
\r {16} {\eightit University of Florida, Gainesville, Florida  32611} \\
\r {17} {\eightit Laboratori Nazionali di Frascati, Istituto Nazionale di Fisica
               Nucleare, I-00044 Frascati, Italy} \\
\r {18} {\eightit University of Geneva, CH-1211 Geneva 4, Switzerland} \\
\r {19} {\eightit Glasgow University, Glasgow G12 8QQ, United Kingdom}\\
\r {20} {\eightit Harvard University, Cambridge, Massachusetts 02138} \\
\r {21} {\eightit The Helsinki Group: Helsinki Institute of Physics; and Division of
High Energy Physics, Department of Physical Sciences, University of Helsinki, FIN-00044, Helsinki, Finland}\\
\r {22} {\eightit Hiroshima University, Higashi-Hiroshima 724, Japan} \\
\r {23} {\eightit University of Illinois, Urbana, Illinois 61801} \\
\r {24} {\eightit The Johns Hopkins University, Baltimore, Maryland 21218} \\
\r {25} {\eightit Institut f\"{u}r Experimentelle Kernphysik, 
Universit\"{a}t Karlsruhe, 76128 Karlsruhe, Germany} \\
\r {26} {\eightit High Energy Accelerator Research Organization (KEK), Tsukuba, 
Ibaraki 305, Japan} \\
\r {27} {\eightit Center for High Energy Physics: Kyungpook National
University, Taegu 702-701; Seoul National University, Seoul 151-742; and
SungKyunKwan University, Suwon 440-746; Korea} \\
\r {28} {\eightit Ernest Orlando Lawrence Berkeley National Laboratory, 
Berkeley, California 94720} \\
\r {29} {\eightit University of Liverpool, Liverpool L69 7ZE, United Kingdom} \\
\r {30} {\eightit University College London, London WC1E 6BT, United Kingdom} \\
\r {31} {\eightit Massachusetts Institute of Technology, Cambridge,
Massachusetts  02139} \\   
\r {32} {\eightit Institute of Particle Physics: McGill University,
Montr\'{e}al, Canada H3A~2T8; and University of Toronto, Toronto, Canada
M5S~1A7} \\
\r {33} {\eightit University of Michigan, Ann Arbor, Michigan 48109} \\
\r {34} {\eightit Michigan State University, East Lansing, Michigan  48824} \\
\r {35} {\eightit Institution for Theoretical and Experimental Physics, ITEP,
Moscow 117259, Russia} \\
\r {36} {\eightit University of New Mexico, Albuquerque, New Mexico 87131} \\
\r {37} {\eightit Northwestern University, Evanston, Illinois  60208} \\
\r {38} {\eightit The Ohio State University, Columbus, Ohio  43210} \\  
\r {39} {\eightit Okayama University, Okayama 700-8530, Japan}\\  
\r {40} {\eightit Osaka City University, Osaka 588, Japan} \\
\r {41} {\eightit University of Oxford, Oxford OX1 3RH, United Kingdom} \\
\r {42} {\eightit University of Padova, Istituto Nazionale di Fisica 
          Nucleare, Sezione di Padova-Trento, I-35131 Padova, Italy} \\
\r {43} {\eightit University of Pennsylvania, Philadelphia, 
        Pennsylvania 19104} \\   
\r {44} {\eightit Istituto Nazionale di Fisica Nucleare, University and Scuola
               Normale Superiore of Pisa, I-56100 Pisa, Italy} \\
\r {45} {\eightit University of Pittsburgh, Pittsburgh, Pennsylvania 15260} \\
\r {46} {\eightit Purdue University, West Lafayette, Indiana 47907} \\
\r {47} {\eightit University of Rochester, Rochester, New York 14627} \\
\r {48} {\eightit The Rockefeller University, New York, New York 10021} \\
\r {49} {\eightit Istituto Nazionale di Fisica Nucleare, Sezione di Roma 1,
University di Roma ``La Sapienza," I-00185 Roma, Italy}\\
\r {50} {\eightit Rutgers University, Piscataway, New Jersey 08855} \\
\r {51} {\eightit Texas A\&M University, College Station, Texas 77843} \\
\r {52} {\eightit Texas Tech University, Lubbock, Texas 79409} \\
\r {53} {\eightit Istituto Nazionale di Fisica Nucleare, University of Trieste/\
Udine, Italy} \\
\r {54} {\eightit University of Tsukuba, Tsukuba, Ibaraki 305, Japan} \\
\r {55} {\eightit Tufts University, Medford, Massachusetts 02155} \\
\r {56} {\eightit Waseda University, Tokyo 169, Japan} \\
\r {57} {\eightit Wayne State University, Detroit, Michigan  48201} \\
\r {58} {\eightit University of Wisconsin, Madison, Wisconsin 53706} \\
\r {59} {\eightit Yale University, New Haven, Connecticut 06520} \\
\end{center}

}   

\date{\today}

\begin{abstract}
We report a measurement of the forward-backward charge asymmetry of electrons
from $W$ boson decays in \ppbar collisions at $\sqrt{s}=1.96$ TeV
using a data sample of $170 \minvpb$ collected by the Collider Detector at
Fermilab.
The asymmetry is measured as a function of electron rapidity
and transverse energy
and provides new input on the momentum fraction dependence of the 
$u$ and $d$ quark parton distribution functions within the proton.
\end{abstract}

\pacs{13.38.Be, 13.85.Qk, 14.60.Cd, 14.70.Fm, }
\keywords{W bosons, leptons, asymmetry}

\maketitle

\section{Introduction} \label{secintroduction}

A necessary input for cross section calculations at a hadron collider
is an estimate of the momentum distribution of the incoming partons that
participate in the hard-scattering process.   The probability of finding 
a parton carrying momentum fraction $x$ within the incoming hadron is 
expressed in the parton distribution function (PDF).  
At the Tevatron, any cross section calculation will have to 
integrate over the proton and anti-proton PDFs.  
Presently, many measurements at the Tevatron have significant 
uncertainties associated with the choice of PDF.  
These uncertainties will become more important as the datasets 
continue to grow. For example, PDF uncertainty is expected 
to be among the dominant systematic uncertainties in a precision 
determination of the $W$ boson mass.

The PDFs are not calculable and must be determined using measurements 
from a wide range of scattering processes~\cite{cteq,mrst}.
Measurement of the forward-backward charge asymmetry in
\ppbar$ \rightarrow W^{\pm}+X$ provides important input on the ratio 
of the $u$ and $d$ quark components of the PDF. 
Since $u$ quarks carry, on average, a higher fraction of the proton
momentum than $d$ quarks~\cite{uoverd}, 
a $W^+$ produced by $u \bar d \rightarrow W^+$ tends to
be boosted forward, in the proton direction.
Similarly, a $W^-$ tends to be boosted backward.
This results in a non-zero forward-backward charge asymmetry defined as
\begin{equation} \label{eqwasymmetry}
A(y_W)=\frac{d\sigma(W^+) / dy_W - d\sigma(W^-) / dy_W}
{d\sigma(W^+) / dy_W +d\sigma(W^-) / dy_W}\,,
\end{equation}
where $y_W$ is the rapidity of the $W$ bosons and 
$d\sigma(W^\pm)/dy_W$ is the differential cross section for
$W^+$ or $W^-$ boson production.

Leptonic decays of the $W$ boson, in our case $W \rightarrow e\nu$, 
provide a
high purity sample for measuring this asymmetry. 
However, because $p_Z$ of the neutrino is unmeasured, 
$y_W$ is not directly determined, and we instead measure 
\begin{equation} \label{eqleptonasymmetry}
A(\eta_e)=\frac{d\sigma(e^+) / d\eta_e - d\sigma(e^-) /
d\eta_e}{d\sigma(e^+) / d\eta_e +d\sigma(e^-) / d\eta_e}\,, 
\end{equation}
where $\eta_e$ is the electron pseudorapidity~\cite{geomnote}.
By assuming the $W \rightarrow e\nu$ decays are described by the Standard
Model $V-A$ couplings, $A(\eta_e)$ probes the PDF.

Previous measurements of the asymmetry~\cite{run1asym},
using $110\minvpb$ of \ppbar data
at $\sqrt{s}=1.8~{\rm TeV}$ 
collected by the Collider Detector at Fermilab (CDF), 
have provided constraints on the 
PDFs for $u$ and $d$ quarks at momentum transfer of $Q^2\approx {M_W}^2$.
In this article we describe a new measurement based on  data 
collected with the CDF II detector at $\sqrt{s}=1.96~{\rm TeV}$ 
corresponding to an integrated luminosity of $170~{\rm pb}^{-1}$. 
We measure the asymmetry in two regions of electron $E_T$ 
that probe different ranges of $y_W$ and thus increase sensitivity to the 
PDFs in the region $x > 0.3$ where currently they are least constrained.

\section{Detector Description} \label{secdetector}

The CDF II detector~\cite{TDR} has undergone a major upgrade 
since the previous data-taking period. 
The components relevant to this measurement are described here.

Tracking detectors immersed within a $1.4~{\rm T}$ solenoidal magnetic field
are used to reconstruct the trajectories (tracks) and measure the momentum of
charged particles. 
The Central Outer Tracker (COT) is a $3.1~{\rm m}$ long open-cell drift chamber
which provides track measurements (hits) in 96 layers in the radial 
range $40~{\rm cm} < r < 137~{\rm cm}$~\cite{COT}.
Closer to the beam, a silicon tracking system~\cite{Silicon} provides precise hits 
from eight layers of sensors spanning $1.3~{\rm cm} < r < 28~{\rm cm}$ and 
extending up to $1.8~{\rm m}$ along the beam line.
The COT allows track reconstruction in the range  $|\eta|\simlt 1$.
The silicon detector extends that range to $|\eta| \simlt 2.5$.

Segmented electromagnetic (EM) and hadronic calorimeters surround the tracking
system and measure the energy of particles~\cite{calors}.
The energy of electrons is measured by lead-scintillator sampling 
calorimeters. 
In the central region, $|\eta|<1.1$, the calorimeters are arranged in a
projective 
barrel geometry and measure EM energy
 with a resolution of $[\sigma(E_T)/E_T]^2 = (13.5\%)^2
 /E_T ({\rm GeV}) + (2\%)^2
 $.
In the forward region, $1.2<|\eta|<3.5$, the calorimeters are arranged in a
projective ``end-plug'' geometry and measure EM energy with
a resolution of $[\sigma(E_T)/E_T]^2 = (14.4\%)^2
/E_T ({\rm GeV}) + (0.7\%)^2
$.

Both central and forward EM calorimeters are instrumented with finely 
segmented detectors which measure shower position
at a depth where energy deposition by a typical shower reaches its 
maximum. 
In the central region we use proportional wire chambers with 
cathode strip readout, 
in the forward region shower position is 
measured by two layers of $5~{\rm mm}$ wide 
scintillating strips with a
stereo angle of 45 degrees between them.

\section{Data Sets and Selection} \label{secdatasets}

Our signal sample is comprised of $W \rightarrow e\nu$ candidate events, 
and a sample of $Z^0 \rightarrow e^+ e^-$ candidate events is
used to calibrate the charge identification.
Events of interest are initially 
selected by an online trigger system
with differing requirements for the central and forward regions.
For $W$ candidates, the central trigger requires 
an EM energy cluster with $E_T > 18~{\rm GeV}$ and a matching track with
$p_T> 9~\mGeVc$.
To avoid any potential charge bias in the track trigger efficiency,
we also accept events from a trigger which requires 
an EM energy cluster with $E_T > 20~{\rm GeV}$ and
missing transverse energy ($\MET$) of at least $25~{\rm GeV}$,
but has no explicit track requirement.
The forward trigger for $W$ candidates requires
an EM energy cluster with $E_T > 20~{\rm GeV}$
and $\MET > 15~{\rm GeV}$.
A backup trigger drops the $\MET$ requirement and
is used to estimate
the QCD jet background contribution.
The trigger for $Z$ candidates requires two EM energy clusters
with $E_T > 18~{\rm GeV}$.

The criteria used to identify the electron and positron candidates,
which are described in detail in reference~\cite{wcsprd} and summarized  
below, 
are designed to reject the energy deposits from photons or QCD jets.
\begin{itemize}
\item   $E_T > 25\,\rm{GeV}$,
\item 	$F_{\rm Iso} <0.1$, 
where $F_{\rm Iso} \equiv$  additional energy in an ``isolation'' 
cone, of angular radius  $R = \sqrt{(\Delta\phi)^2 + (\Delta\eta)^2} = 0.4$
centered on the electron, divided by the electron energy,
\item  A small amount of associated hadronic energy, less than $5\%
$
of the EM energy,
\item  The shower shape in the EM calorimeter and shower maximum
detector must be consistent with that observed from test-beam data,
\item The position along the beamline of the \ppbar collision, $z_0$,
is well reconstructed and $|z_0|<60~{\rm cm}$~\cite{zvtx},
\item  A track consistent with the position and energy 
measured in the calorimeter.
\end{itemize}

COT tracks, reconstructed independent of the calorimeter measurement,
can be compared to it in position and momentum.
However, the coverage of the COT is limited to $|\eta|\simlt 1$.
To extend the measurement to higher $|\eta|$, 
we instead use silicon tracks
reconstructed by a new calorimeter-seeded algorithm as
described below.
Two points and a signed curvature define a unique helix. 
The positions of the electromagnetic shower and of the \ppbar collision provide
the two points.
The curvature of the trajectory is predicted from the transverse
energy measured by the calorimeter.
These two points and the curvature are used to generate two
seed helices and associated covariance matrices, one for each charge
hypothesis. 
Those seed helices are then projected into the silicon detector
where hits are attached using a road-based search and
requiring at least 4 attached hits with $\chi^2/{\rm dof}<8$.
If silicon tracks are fit for both charge hypotheses,
the $\chi^2/{\rm dof}$ is used to identify the charge with the best fit,
and cases with $\Delta\chi^2/{\rm dof} < 0.5$ are rejected as ambiguous.

The relative alignment of the silicon detector and 
the calorimeter is determined using a sample of 
well identified $e^\pm$ with both COT and silicon tracks.
To avoid a charge bias from the $W$ charge asymmetry, 
we explicitly equalize the number of events of each charge
used in the alignment
for $\eta>0$ and separately for $\eta<0$.
Offsets of $\cal{O}$$(1~{\rm mm})$ 
and rotations of $\cal{O}$$(10~{\rm mrad})$ are measured and corrected.
The resulting position resolution in the forward calorimeter is 
measured to be $1~{\rm mrad}$, consistent with the design expectation.

Candidate $W \to e \nu$  events are required to have exactly one such
$e^\pm$ candidate as well as 
$\MET > 25\,\rm{GeV}$ and transverse mass in the range 
$50\,\mGeVcc < M_T<100\,\mGeVcc$.
To suppress backgrounds from QCD and Drell-Yan processes, 
we require that there be
no other EM energy depositions with $E_T > 25\,\rm{GeV}$.
The selected sample contains  49,124 central and 28,806 forward events.

\section{Measurement of the Charge Asymmetry} \label{secwa}

\begin{figure}[h]
\includegraphics[width = 15cm]{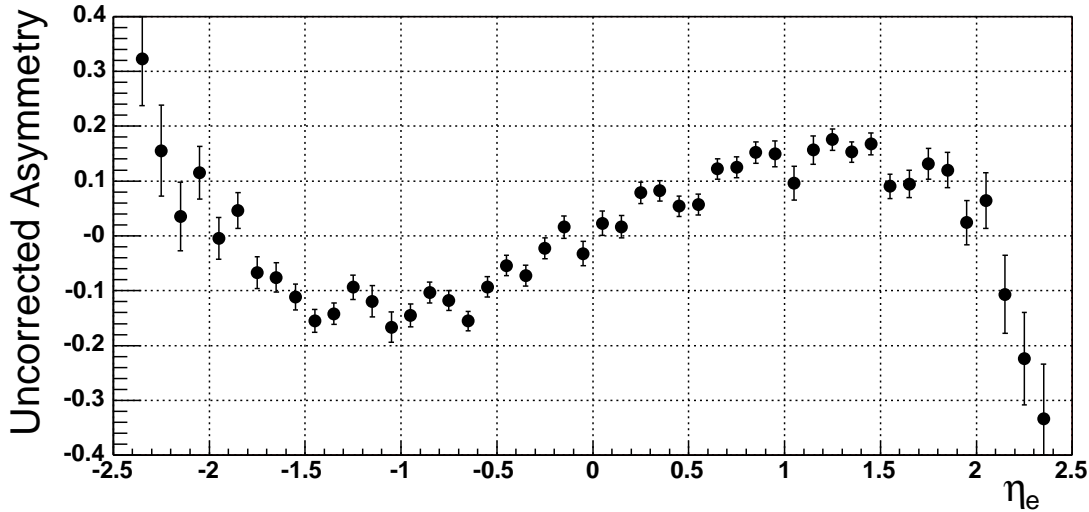}
\caption{
The raw, uncorrected, charge asymmetry is plotted as a function of
electron $\eta$. 
}
\label{figrawasym}
\end{figure}

Directly measured in the experiment and shown in Figure~\ref{figrawasym}
is the raw, uncorrected, asymmetry.  
In order to reconstruct $A(\eta_e)$,
the measurement needs to be corrected 
for the effects of charge misidentification and background contributions.
These $\eta$ dependent corrections are applied bin-by-bin, 
and binning coarser than shown in Figure \ref{figrawasym}  is used 
to reduce the effect of the uncertainty from these corrections.

\subsection{Charge Misidentification}
\label{seccfr}

The electron identification is constructed, and observed, to have a
charge symmetric efficiency.
However, resolution effects can lead to misidentification of the charge,
which dilutes the asymmetry.
Residual misalignments in the silicon detector and calorimeters
could give rise to a bias in the charge identification that would
directly bias the asymmetry. 
We measure the probability of such misidentification and correct for it.
Calling that probability 
$f_+$ for $e^+$ and $f_-$ for $e^-$, 
the corrected asymmetry can be computed from the raw asymmetry as 
$A = (A_{\rm raw} + f_+ - f_-)/(1 - f_+ - f_-)$.

\begin{figure}[h]
\mbox{
\includegraphics[width=15cm]{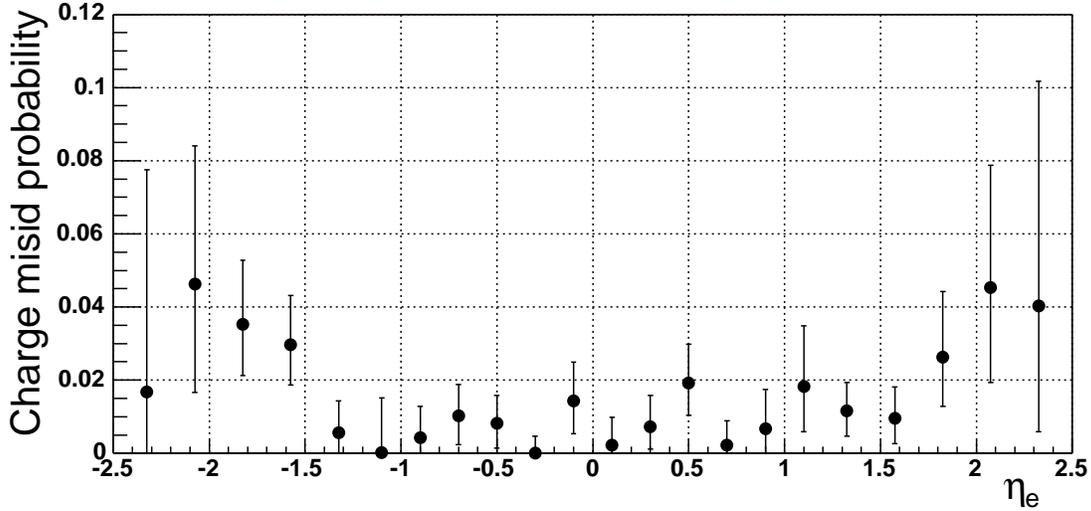}
}
\caption{
The charge misidentification probability is plotted as a function of
electron $\eta$. 
}
\label{figcfr}
\end{figure}

\begin{figure}[h]
\mbox{
\includegraphics[width=15cm]{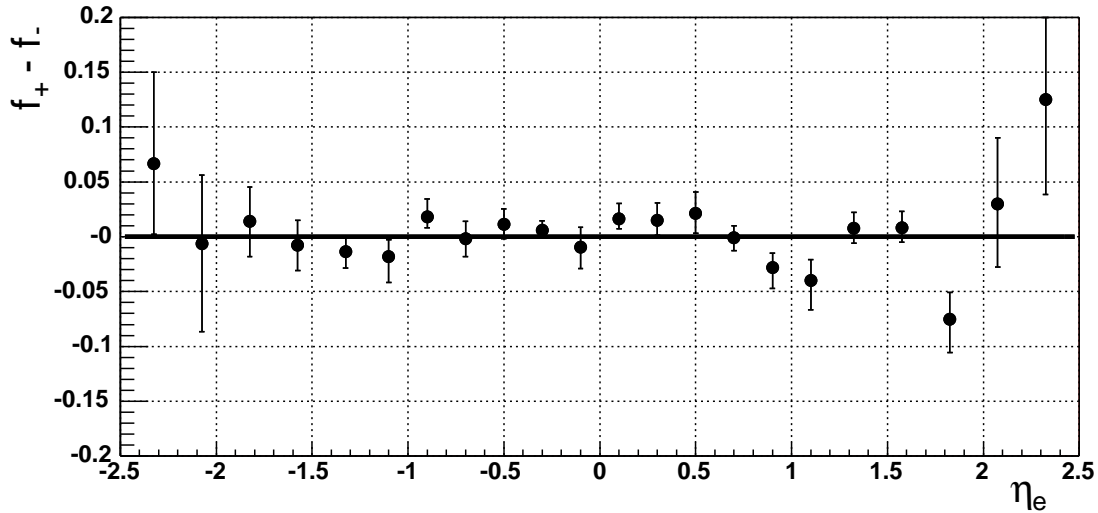}
}
\caption{
The difference in charge misidentification probability of $e^+$ and $e^-$, 
$f_+(\eta) - f_-(\eta)$, 
is plotted as a function of electron $\eta$. 
}
\label{figcfr2}
\end{figure}

We measure $f_\pm(\eta)$ with $Z^0 \rightarrow e^+ e^-$ events
where a track matched to one lepton tags the charge of the other.
The tagging leg must have $|\eta|<1.5$, and COT track information is used if
it is available. 
The average misidentification probability i.e., without distinguishing between
$e^+$ and $e^-$, is
shown as a function of $\eta$ in Figure~\ref{figcfr}.
The difference between the misidentification probability for $e^+$ and $e^-$
 is shown in Figure~\ref{figcfr2}, 

\subsection{Background Corrections} \label{secbkgd}

We correct the measurement for the contributions
of three sources of background: 
QCD jets, $Z^0\rightarrow e^+e^-$, and 
$W \rightarrow \tau\nu \rightarrow e \nu \nu \nu$.

The background contribution from QCD jets faking the $W \rightarrow e\nu$ 
signature is measured by comparing the isolation of the $e^\pm$ candidate 
to the $\MET$ in the event~\cite{wcsprd}.
Electrons from $W$ decays tend to be isolated i.e., 
have low $F_{\rm iso}$ values,
while background from QCD jets have larger values.
Similarly, $W \rightarrow e\nu$ events have large $\MET$
while QCD jets have lower values.
If there is no correlation between isolation and $\MET$ for QCD jets, 
we can measure their shapes in the non-$W$ regions and extrapolate them 
into the signal region.
Studies of these variables demonstrate that they are
not correlated if the selection requirements related to
the EM shower shape are relaxed.
Including those requirements suppresses 
events with high values of $F_{\rm iso}$, 
which makes the extrapolation
statistically imprecise and degrades our ability to measure any potential 
correlation, so we remove them in estimating the QCD jet background.
That results in an overestimate of the background, 
but it yields a statistically and systematically robust estimate.
This measured upper bound on the background fraction
is $2\%
$ for $|\eta|<1$
and increases to about $15\%
$ for $|\eta|>2$.
We correct the raw asymmetry by a factor of $1 + F_{\rm QCD}$,
where for the background fraction, $F_{\rm QCD}$,
we use half the measured upper bound, with 
uncertainties of $\pm 50\%
$. 
Since we have only an upper limit, 
this choice provides full coverage of the actual value at $2\sigma$.

$Z^0\rightarrow e^+e^-$ events in which one of the leptons is lost 
represent a small, but asymmetric background~\cite{zasym}.
This background contribution is determined with a Monte
Carlo calculation using the {\sc PYTHIA} generator\cite{pythia},
and it corresponds to about $1\%
$ of the signal.
$W \rightarrow \tau\nu \rightarrow e\nu\nu\nu$ events 
bias the measured asymmetry because the
$\tau$ decay dilutes the information available in the $e^\pm$ direction.
This background 
contribution is about $4\%
$ of the signal.
The number of $e^+$ and $e^-$ events predicted for these backgrounds
are subtracted from the measured values bin-by-bin in $\eta$.

\begin{figure}[h]
\includegraphics[width=15cm]{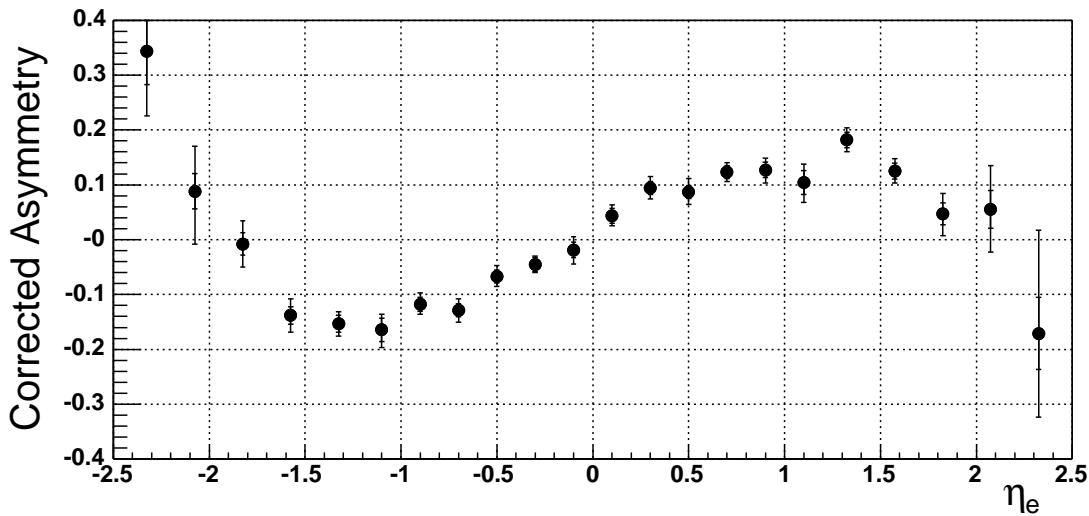}
\caption{
The fully corrected charge asymmetry is plotted as a function of
electron $\eta$. 
Both statistical and total (statistical+systematic) uncertainties are shown.
}
\label{figcorrasym}
\end{figure}

Figure \ref{figcorrasym} shows the fully corrected $A(\eta_e)$.

\subsection{$E_T$ dependence}

The asymmetry probes a large range of $x$ for the parent
$u$ and $d$ quarks,
from an upper value of approximately $0.5$, where 
valence quarks dominate,
down to $2 \times 10^{-3}$, where sea quarks dominate.
Large values of $y_W$ correspond to the extreme values of $x$.
For example, 
a high-$x$ $u$ quark and a low-$x$ $\bar d$ quark lead to $W^+$ with large 
$p_Z$ and therefore large $y_W$.
The $V-A$ couplings in the $W\rightarrow e\nu$ decay 
cause the $e^+$ to be preferentially emitted opposite the $W^+$ flight
direction.
The electron asymmetry, $A(\eta_e)$, is
a convolution of these competing production and decay asymmetries,
which results in the sign change of $A(\eta_e)$ at large $|\eta_e|$.

Direct sensitivity to the PDF would be improved by
reducing the decay asymmetry effect 
e.g., by reconstructing the $W$ direction.
The unmeasured $p_Z$ of the neutrino and the poor $\MET$ resolution 
complicate this reconstruction.
However, we can improve the correspondence between
$\eta_e$ and $y_W$ based on the kinematics of just the electron,
which is well measured.
The neutrino $p_Z$ ambiguity is a smaller effect for
electrons with high $E_T$ than for those at low $E_T$.
We exploit this by separating the asymmetry measurement into bins of electron
$E_T$. 
The size of the statistical and systematic uncertainties allow two bins,
$25~{\rm GeV} < E_T < 35~{\rm GeV}$ and
$35~{\rm GeV} < E_T < 45~{\rm GeV}$.
For a given $\eta_e$, the two $E_T$ regions probe different ranges of $y_W$,
and therefore $x$,
and the higher $E_T$ bin corresponds to a narrower range.
As a result, measuring the asymmetry separately in the two bins 
allows a finer probe of the $x$ dependence.


\subsection{Systematic uncertainties}

The corrections for charge misidentification and background contributions
are measured and applied separately for each $E_T$ bin since they are 
$E_T$ dependent. 
The statistical uncertainty on the charge misidentification correction
dominates the systematic uncertainty on the asymmetry measurement.
The uncertainty from the QCD jet background correction is small, 
and the other background uncertainties are negligible.

Detector misalignments can induce an inherent charge bias.
Such biases would be naturally corrected by the charge
misidentification probabilities measured from the data.
Nonetheless, we check the robustness of the charge determination by
varying the alignment corrections within their uncertainties and 
verifying that the resulting changes in the asymmetry are not significant.
We also verify that using COT tracks, when they are available, instead of
silicon tracks results in no significant difference.

CP invariance requires $A(-\eta_e) = -A(\eta_e)$.
The fully corrected data shown in Figure \ref{figcorrasym}
show no evidence of CP asymmetry, 
the level of agreement is characterized by $\chi^2/{\rm dof} =9.5/11$.
The $\pm\eta_e$ data are folded together 
to obtain a more precise measure of A($|\eta_e|$).

These results are most useful as input to future global PDF fits.
Such fits use Monte Carlo generators without a full detector
simulation.
We have studied possible biases introduced by detector effects by comparing
the asymmetry from a {\sc PYTHIA} Monte Carlo generator 
to the fully simulated results and found no significant effects.


\subsection{Results}

\begin{figure}[t]
\includegraphics[width = 15cm]{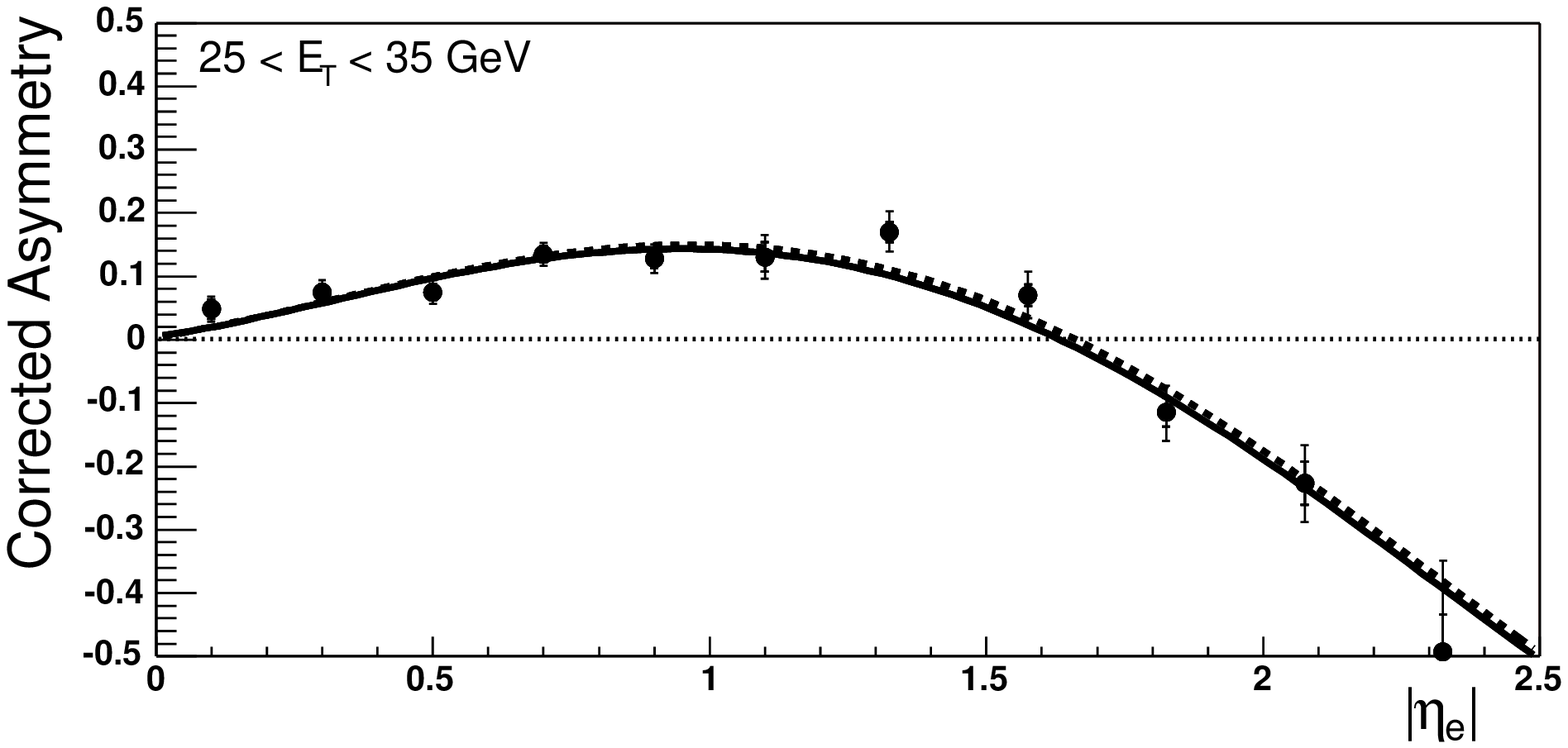}
\includegraphics[width = 15cm]{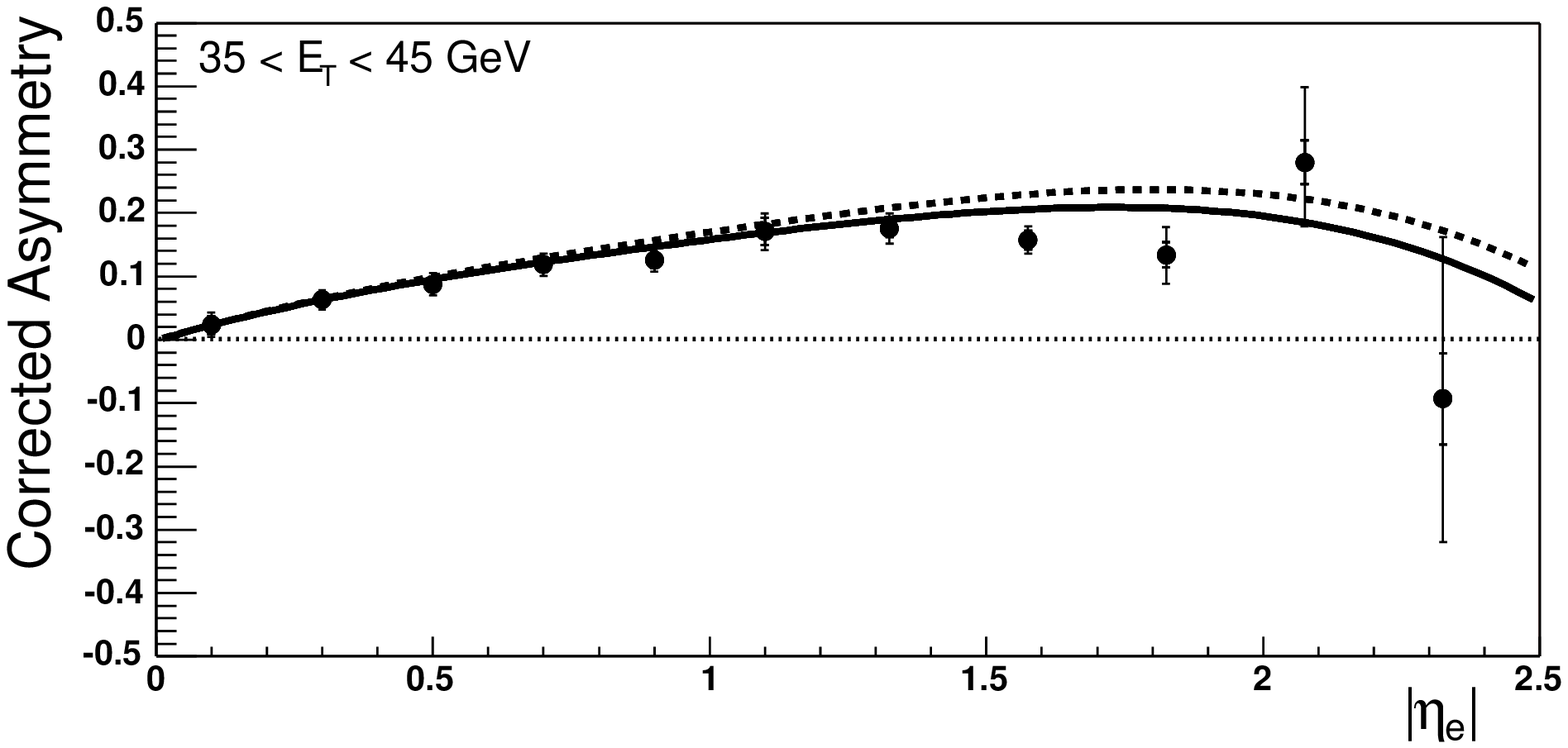}
\caption{
The measured asymmetry, $A(|\eta_e|)$, is plotted and
predictions from the CTEQ6.1M (solid) and MRST02 (dashed) PDFs
are compared using a NLO RESBOS calculation.
Both statistical and total (statistical+systematic) uncertainties are shown.
The upper plot is for $25<E_T<35$ GeV.
The lower plot is for $35<E_T<45$ GeV.
}
\label{figpdfcompareET}
\end{figure}

The measured asymmetry, $A(|\eta_e|)$, 
is listed in Table~\ref{tabresults} and
plotted in Figure \ref{figpdfcompareET} for the two $E_T$ regions.
Predictions 
from CTEQ~\cite{cteq} and MRST~\cite{mrst} PDFs,
which fit to previous CDF results~\cite{run1asym}, are shown for comparison.
Those predictions use a NLO RESBOS Monte Carlo calculation 
with soft gluon resummation to model the $W$ $p_T$ distribution,
to which they can be sensitive~\cite{RESBOS}.
Since the previous measurements upon which these predictions are
based are least constraining for $|\eta|>1$ 
and do not separate the $E_T$ dependence,
inclusion of our results will further constrain future fits and improve the
predictions.

\begin{table}[t!]
\begin{ruledtabular}
\begin{tabular}{cccc}
  $|\eta_e|$&
 \multicolumn{3}{c}{$A(|\eta_e|)$} \\ \hline
   &
  $E_T>25$ &
 $25 < E_T < 35$ &
 $35 < E_T < 45$ \\
\hline
  $0.11$ & 
  ${3.4}~^{+1.6}_{-1.5}$&
  ${4.8}~\pm~{2.0}$ & 
  ${2.3}~\pm~{1.9}$ \\
  $0.30$ & 
  ${6.2}~\pm~{1.2}$ &
  ${7.5}~\pm~{1.9}$ & 
  ${6.3}~\pm~{1.5}$ \\
  $0.50$ & 
  ${7.5}~\pm~{1.5}$ &
  ${7.5}~\pm~{1.9}$ & 
  ${8.8}~\pm~{1.8}$ \\
  $0.70$ & 
  ${12.6}~\pm~{1.3}$ &
  ${13.5}~\pm~{1.8}$ & 
  ${11.8}~\pm~{1.7}$ \\
  $0.89$ & 
  ${12.2}~^{+1.6}_{-1.4}$ &
  ${12.8}~\pm~{2.3}$ & 
  ${12.6}~^{+1.7}_{-1.9}$ \\
  $1.09$ & 
  $13.8~\pm~2.3$ &
  ${13.1}~\pm~{3.5}$ & 
  ${17.1}~\pm~{2.9}$ \\
  $1.33$ & 
  ${16.8}~\pm~1.6$ &
  ${17.0}~^{+3.4}_{-3.0}$ & 
  ${17.6}~\pm~{2.4}$ \\
  $1.57$ & 
  $13.0~\pm~1.8$ &
  ${7.0}~^{+3.8}_{-3.6}$ & 
  ${15.7}~\pm~{2.2}$ \\
  $1.81$ & 
  $2.9~\pm~2.9$ &
  ${-11.5}~^{+4.2}_{-4.5}$ & 
  ${13.4}~^{+4.4}_{-4.6}$ \\
  $2.04$ & 
  ${-0.4}~^{+6.2}_{-5.7} $ &
  ${-23}~\pm~{6}$ & 
  ${28}~^{+12}_{-10}$ \\
  $2.31$ & 
  ${-29}~\pm~{10}$ &
  ${-49}~\pm~{14}$ & 
  ${-9}~^{+26}_{-23}$ 
\end{tabular}
\end{ruledtabular}
\caption{\label{tabresults} The measured asymmetry values are tabulated in
percent with combined statistical and systematic uncertainties. 
The listed $|\eta_e|$ is the event weighted average.
Asymmetric uncertainties listed for some values arise because of the
Poisson and binomial statistics inherent in the event counting.
}
\end{table}

\acknowledgments 
\label{secacknowledgments}

  We thank the Fermilab staff and the technical staffs of the participating institutions for their vital
  contributions. 
  We thank Pavel Nadolsky for providing the RESBOS predictions.
We thank the Fermilab staff and the technical staffs of the participating institutions for their vital contributions. This work was supported by the U.S. Department of Energy and National Science Foundation; the Italian Istituto Nazionale di Fisica Nucleare; the Ministry of Education, Culture, Sports, Science and Technology of Japan; the Natural Sciences and Engineering Research Council of Canada; the National Science Council of the Republic of China; the Swiss National Science Foundation; the A.P. Sloan Foundation; the Bundesministerium fuer Bildung und Forschung, Germany; the Korean Science and Engineering Foundation and the Korean Research Foundation; the Particle Physics and Astronomy Research Council and the Royal Society, UK; the Russian Foundation for Basic Research; the Comision Interministerial de Ciencia y Tecnologia, Spain; in part by the European Community's Human Potential Programme under contract HPRN-CT-2002-00292; and the Academy of Finland.
\bibliography{waprd}

\begin{thebibliography}{7}
\bibitem[{geo()}]{geomnote}
\bibinfo{note}{We use a cylindrical coordinate system about the beam axis in
  which $z$ is the proton direction, $\theta$ is the polar angle, $\phi$ is the
  azimuthal angle, and pseudorapidity $\eta \equiv -\ln \tan (\theta /2)$
  is an excellent approximation for the rapidity of low mass particles. 
  $E_T \equiv E \sin
  \theta$ and $p_T \equiv p \sin \theta$ where $E$ is energy measured by the
  calorimeter and $p$ is momentum measured by the spectrometer. The transverse
  momentum of the neutrino can be inferred from the missing transverse energy,
  \MET$\equiv -\sum_i E_T^i {\bf n}_i$, where ${\bf n}_i$ is the unit vector in
  the azimuthal plane that points from the beamline to the center of the 
  $i$th calorimeter tower. The transverse mass, $M_T$, measures the invariant
  mass using only momentum and energy transverse to the beam direction.}
\bibitem[{zvt()}]{zvtx}
\bibinfo{note}{The $z_0$ position is found by vertexing all
   the other tracks from the $p \bar p$ interaction.
   If more than one interaction point is found, we use the one with the 
   largest associated $p_T$, estimated as the scalar 
   $\Sigma p_T$ of all tracks used in the vertex fit.}
\end{thebibliography}

\end{document}